\documentclass[a4paper]{article}
\usepackage{multirow}
\usepackage{CJK}
\usepackage{INTERSPEECH2019}

\title{HI-MIA : A FAR-FIELD TEXT-DEPENDENT SPEAKER VERIFICATION DATABASE AND THE BASELINES}

 \name{Xiaoyi Qin$^{1}$ \qquad Hui Bu$^{2}$ \qquad Ming Li$^{1,3}$}

\address{$^{1}$ Data Science Research Center, Duke Kunshan University, Kunshan, China \\
			    $^{2}$ Beijing Shell Shell Technology Co. Ltd., Beijing, China \\
			    $^{3}$ School of Computer Science, Wuhan University, Wuhan, China
			    }
\email{}
\begin{document}

\maketitle

\begin{abstract}
This paper presents a far-field text-dependent speaker verification database named \textit{HI-MIA}. We aim to meet the data requirement for far-field microphone array based speaker verification since most of the publicly available databases are single channel close-talking and text-independent. The database contains recordings of 340 people in rooms designed for the far-field scenario. Recordings are captured by multiple microphone arrays located in different directions and distance to the speaker and a high-fidelity close-talking microphone. Besides, we propose a set of end-to-end neural network based baseline systems that adopt single-channel data for training. Moreover, we propose a testing background aware enrollment augmentation strategy to further enhance the performance. Results show that the fusion systems could achieve 3.29\% EER in the far-field enrollment far field testing task and 4.02\% EER in the close-talking enrollment and far-field testing task.

\end{abstract}


\section{Introduction}

The goal of speaker verification is to verify whether the testing audio is indeed  uttered by the target speaker. Recently, many open and free speech databases with thousands of speakers become publicly available. Most of the databases (e.g. AISHELL2\cite{aishell2_2018}, Librispeech\cite{librispeech}, Voxceleb1\&2 \cite{nagrani_voxceleb:_2017,chung_voxceleb2_2018} ) are recorded in a close-talking environment without noise. Nevertheless, this recording environment does not match with the far-field scenarios in real world smart home or Internet of Things applications. Speaker verification under noisy and reverberation conditions is one of the challenging topics. The performance of speaker verification systems degrades significantly in the far-field condition where the speech is recorded in an unknown direction and distance (usually between 1m-10m). This problem also occurs in speech recognition. Although we have simulation toolkits to convert the close-talking speech to simulated far-field speech, there still exists significant channel mismatch comparing to the real recordings. Moreover, the goal of the front-end processing methods are different in speaker verification and speech recognition. Therefore, it is essential to develop an open and publicly available far-field multi-channel speaker verification database.

Various approaches considering the single-channel microphone or multi-channel microphone array have been proposed to reduce the impact of the reverberation and environmental noise. Those approaches address the problem at different levels of the text-independent automatic speaker verification (ASV) system. At the signal level, linear prediction inverse modulation transfer function \cite{borgstrom_linear_2012} and weighted prediction error (WPE) \cite{mosner_dereverberation_2018,yoshioka_generalization_2012} methods are used for dereverberation. Deep neural network (DNN) based denoising methods for single-channel speech enhancement \cite{zhao_robust_2014,kolboek_speech_2016,oo_dnn-based_2016,eskimez_front-end_2018} and beamforming for multi-channel speech enhancement \cite{mosner_dereverberation_2018,heymann_neural_2016,warsitz_blind_2007} are explored for ASV system under complex environments. At the feature level, sub-band Hilbert envelopes based features \cite{falk_modulation_2010,sadjadi_blind_2014,ganapathy_feature_2011}, warped minimum variance distortionless response (MVDR) cepstral coefficients \cite{jin_speaker_2010}, power-normalized cepstral coefficients (PNCC) \cite{PNCC} and DNN bottleneck features \cite{yamada_improvement_2013} have been applied to ASV system to suppress the adverse impacts of reverberation and noise. At the model level, reverberation matching with multi-condition training models has achieved good performance.  

Deep learning promotes the application of speaker verification technology greatly. The recognition system has been significantly improved from the traditional i-vector method  \cite{dehak_front-end_2011}to the DNN-based x-vector method\cite{snyder_x-vectors:_2018}. Recently, CNN-based neural networks\cite{cai_exploring_2018} also perform well in the speaker verification task. However, both traditional methods and deep learning approaches are data-driven methods that need large amounts of training data. The lack of real world collected microphone array based far field data limits the development and application of far field speaker verification technology in different scenarios.

In this paper, we introduce a database named \textit{HI-MIA} containing recordings of wake-up words under the smart home scenario. This database covers 340 speakers and a wide range of channels from close-talking microphones to multiple far-field microphone arrays. It can be used for far-field wake-up word recognition, far-field speaker verification and speech enhancement. In addition, we provide a set of baseline systems\cite{xiaoyi_farfield} that are trained with the far-field speaker verification data in the transfer learning manner. With the model pre-trained by a large scale simulated far-field data, the system performs well on both far-field enrollment with far-field testing and close-talking enrollment with far-field testing tasks. With the help of enrollment data augmentation, the performance of close-talking enrollment have been further improved.

\section{The \textit{HI-MIA} database}

\label{sec:data}

\textit{HI-MIA} includes two sub databases,  which are the \textit{AISHELL-wakeup}\footnote{\texttt{http://www.aishelltech.com/wakeup\_data}} with utterances from 254 speakers and the \textit{AISHELL-2019B-eval} dataset with utterances from 86 speakers. \footnote{\texttt{http://www.aishelltech.com/aishell\_2019B\_eval}}%\footnote{\texttt{http://openslr.org/85/} is AISHELL Speaker Verification Challenge 2019 database which subsamples from \textit{HI-MIA}, only mandarin data is include. The \textit{SLR 85 HI-MIA} dataset is a subset of the HI-MIA database in this paper.}
\footnote{\texttt{http://openslr.org/85/} is the AISHELL Speaker Verification Challenge 2019 database which contains one close-talking mic, and 1m,3m and 5m  distance microphone arrays right in front of speaker, only mandarin data is include. The \textit{SLR 85 HI-MIA} dataset is a subset of the HI-MIA database in this paper.}
%The recording condition of each speaker is shown in Table \ref{tab:t1}.
\subsection{AISHELL-wakeup}

\begin{figure}[t]
  \centering
  \includegraphics[width=0.9\linewidth]{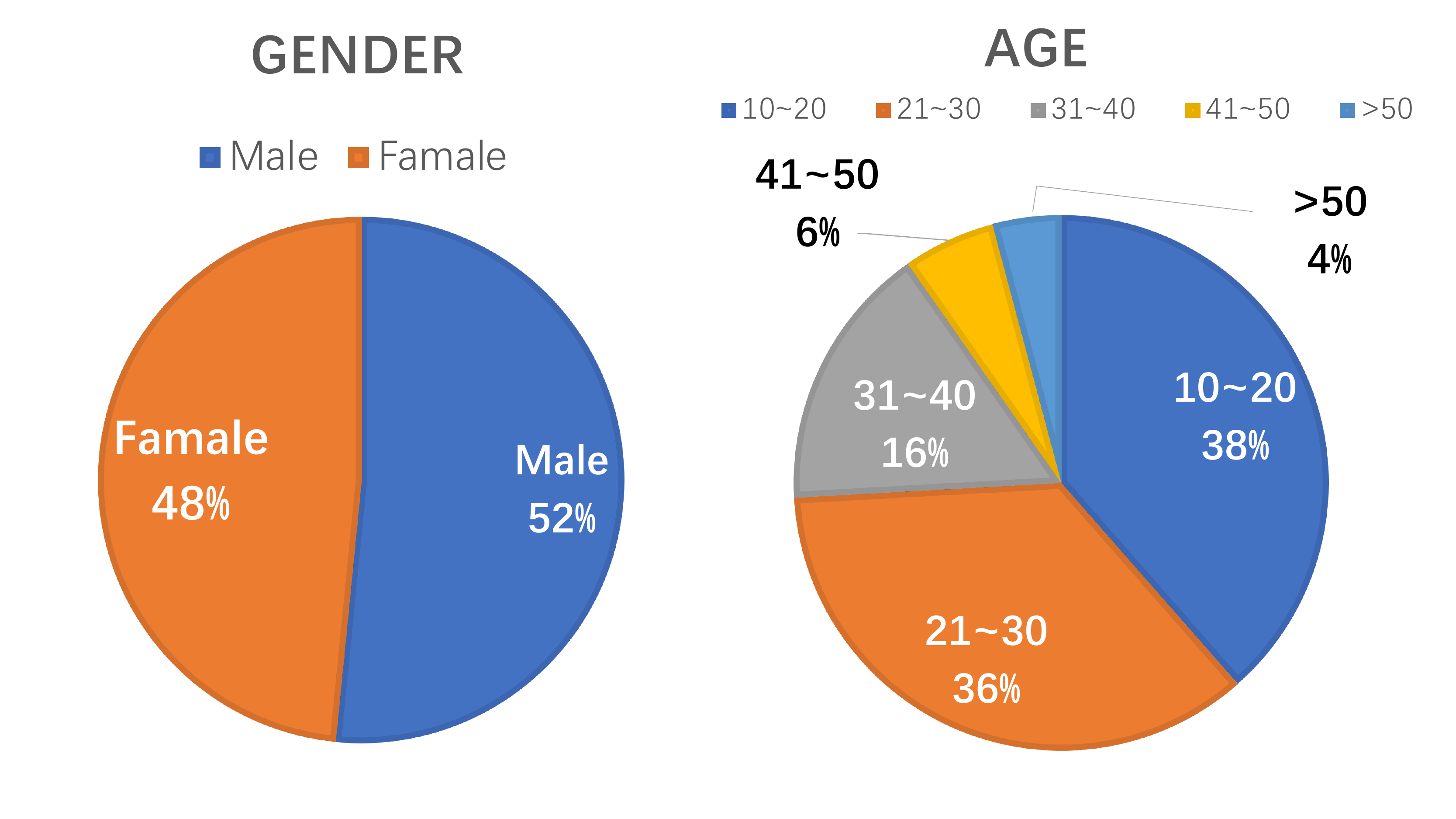}
  \caption{Gender and age distribution}
  \label{fig:gender}
\end{figure}

The \textit{AISHELL-wakeup} database has 3,936,003 utterances with 1,561.12 hours in total. The content of utterances covers two wake-up words, 'ni hao, mi ya ("\begin{CJK}{UTF8}{gbsn}你好，米雅\end{CJK}")' in chinese and 'Hi, Mia' in English. The average duration is around 1 second. The dataset is fairly gender-balanced, with 131 male speakers and 123 female speakers. The distribution of age and gender is shown in Figure \ref{fig:gender}. During the recording process, seven recording devices (one close-talking microphone and six 16-channel circular microphone arrays) were set in a real smart home environment. The duration of utterances recorded by each microphone is 16 hours. The 16-channel circular microphone array records in 16kHz, 16 bit. The close-talking microphone records high fidelity (HiFi) clean speech  in 44.1kHz, 16 bit.

Each speaker recorded 160 utterances, with 120 utterances recorded in a noisy environment and the remaining utterances recorded in the home environment. The details of the database are shown in Table \ref{tab:data details}.

\begin{table*}[t]
  \caption{The details of utterances for each recording speaker}
  \label{tab:data details}
  \centering
  \begin{tabular}[c]{llllll}
    \toprule
    \multicolumn{3}{c}{\textbf{}}&\textbf{AISHELL-wakeup} & \textbf{AISHELL-2019B-eval} \\
    \midrule
  Text ID&  Content & Speed & Environment & Environment\\
    \midrule
  001-020&  ni hao, mi ya & Normal &  \multirow{6}*{TV / Music} & \multirow{6}*{Clean}\\
  021-040& hi,mia & Normal & & \\
  041-060& ni hao, mi ya & Fast& &  \\
  061-080& hi,mia & Fast & & & \\
  081-100& ni hao, mi ya & Slow & & \\
  101-120& hi,mia & Slow &&\\
  \midrule
  121-140& ni hao, mi ya & Normal & \multirow{2}*{Clean} &  \multirow{2}*{TV / Music} &  \\
  141-160& hi,mia & Normal & & & \\
  \bottomrule
  \end{tabular}
\end{table*}
 
\begin{figure}[t]
  \centering
  \includegraphics[width=0.9\linewidth]{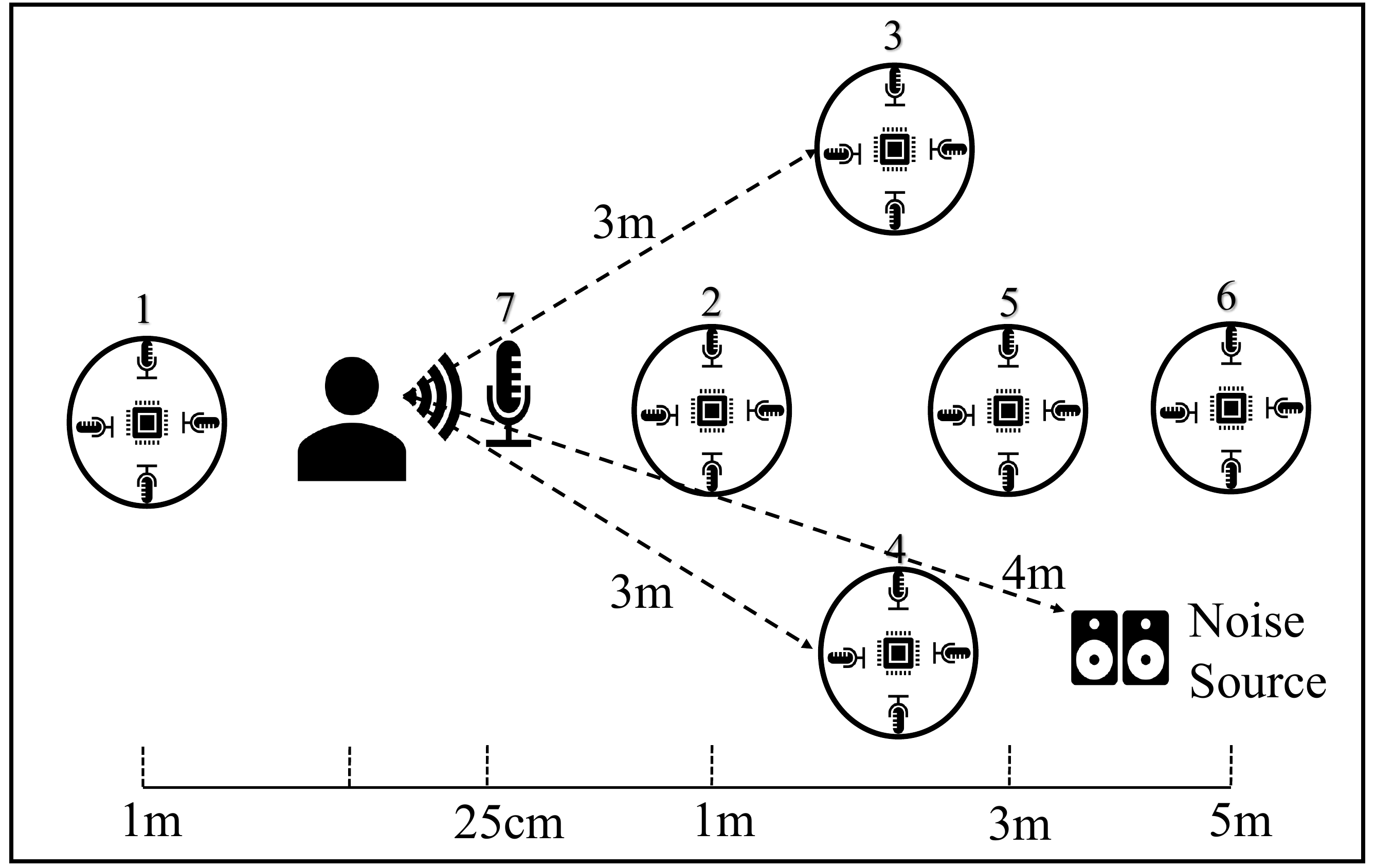}
  \caption{The  setup of the recording environment}
  \label{fig:f1}
\end{figure}

The recordings of each speaker could be cataloged into three subsets according to the speaking speed, i.e.,normal, fast, and slow. We simulated real smart home scenes by adding noise sources such as TV, music, and background noises to the room. The room setting is shown in Figure \ref{fig:f1}. The high-fidelity microphone is 25 cm away from the speaker. The circular microphone arrays are placed around the person with a distance including 1m, 3m and 5m from the person. The noise source is randomly placed close to one of the microphone arrays for each speaker.

\subsection{AISHELL-2019B-eval}

The details of the \textit{AISHELL-2019B-eval} are also shown in table \ref{tab:data details}. The dataset contains recordings of 44 male speakers and 42 female speakers. Different from the \textit{AISHELL-wakeup}, each speaker records 160 utterances, with 120 utterances recorded in a quiet environment and the remaining utterances recorded in the noisy environment. The room setting of \textit{AISHELL-2019B-eval} is the same as the room setting of \textit{AISHELL-wakeup}. We place the noise source in a fixed location four meters away from the speaker.

\section{The Baseline Methods}

\subsection{Deep speaker embedding system}

\subsubsection{Model architecture}

The superiority of deep speaker embedding systems has been shown in text-independent speaker recognition for closed talking \cite{snyder_x-vectors:_2018, cai_exploring_2018} and far-field scenarios \cite{nandwana_robust_2018, dku-voices}. In this paper, we adopt the deep speaker embedding system, which is initially designed for the text-independent speaker verification, as baseline.

The single-channel network structure is the same as in \cite{cai_exploring_2018}. There are three main components in this framework. The first component is a deep CNN (Convolutional neural network) structure based on the well known ResNet-34 architecture (Residual Convolutional Neural Network)\cite{He2016Deep}. We increase the widths (number of channels) of the residual blocks from \{16, 32, 64, 128\} to \{32, 64, 128, 256\}. Then a global statistics pooling (GSP) layer is placed as the encoding layer after the ResNet34, which transforms the feature maps into a fixed-dimensional utterance-level representation. The output of GSP is normalized by its mean and standard deviation. A fully-connected layer then processes the utterance-level representation following by a classification output layer. We add a dropout with a rate of 0.5 before the output layer to prevent over-fitting. Each unit in the output layer refers to a target speaker. The cross-entropy loss is adopted here for measuring the identification error of the training set. 

The network is trained using standard stochastic gradient descent (SGD) with momentum 0.9 and weight decay 1e-4. We use ReduceLROnPlateau in Pytorch to adjust the learning rate, and the initial value is set to 0.01. For each training step, an integer $L$ within $[200,300]$ interval is randomly generated, and each data in the mini-batch is cropped or extended to $L$ frames.

After training, the utterance-level speaker embedding is extracted after the penultimate layer of the neural network for a given utterance. Cosine similarity and PLDA serve as back-end scoring methodes during testing.

%For the multi-channel network model, we use the network model structure described in \cite{cai_3dresnet}. Replace 2d convolution layers in a single channel network with 3d convolution layers. \textit{ResNet2d} refers to the single-channel model and \textit{ResNet3d} refers to the multi-channel model in this paper.

\subsubsection{Training data augmentation for far-field ASV}
Data augmentation can effectively improve the robustness of the deep speaker embedding model\cite{xiaoyi_farfield}. Therefore, we augment the training data with reverberation and noise to simulate far-field speech in real environments and this will reduce the mismatch between training data and test data.

We use the same method as in \cite{xiaoyi_farfield} for data augmentation. \textit{Pyroomacoustics} \cite{pyroomacoustics} is used to simulate far-field recordings by randomly setting the size of the room and arbitrarily locating the the microphone and noise source. The noise source is from MUSAN dataset\cite{musan}, the signal-to-noise ratio (SNR) is 0-20dB. %Besides, we set up a 6-channel microphone array for recording data which matches with the input of ResNet3d model.

\subsection{Model Fine-tuning}
Since we only have limited text-dependent far-field speaker data, if we perform training on these data directly, the text-dependent deep speaker embedding model cannot learn the discriminative speaker information very well, and the model is likely to overfit on test data. Therefore, it is important to use a large amount of text-independent speech data to train a baseline speaker model first.

Therefore we adopt the transfer learning strategy by adapting a text-independent deep speaker embedding model to a text-dependent model. With transfer learning, the adapted text-dependent model takes the advantages of the pre-trained model with a large number of speakers without training the whole network from scratch. After the text-independent deep speaker model is trained, transfer learning adapts the front-end local pattern extractor, the encoding layer and the embedding extraction layer to the text-dependent task. 

\begin{figure}[t]
  \centering
  \includegraphics[width=\linewidth]{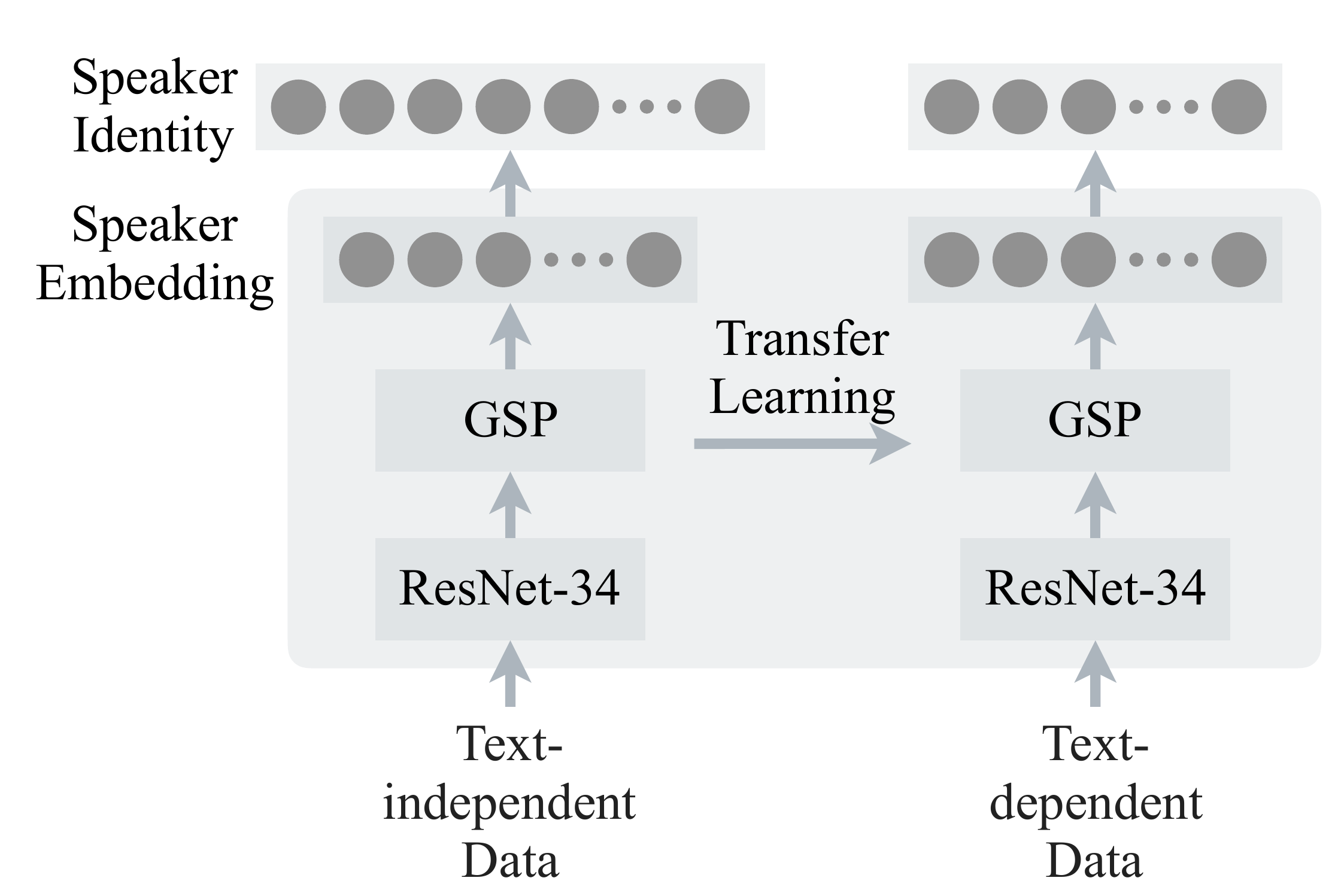}
  \caption{Transfer the text-independent deep speaker embedding model to text-dependent model.}
  \label{fig:fine-tune}
\end{figure}

Figure \ref{fig:fine-tune} shows the transfer learning process of the text-dependent deep speaker embedding model.

\subsection{Enrollment data augmentation}

In the close-talking enrollment with far-field testing task, the mismatch between the enrollment and testing data degrades the performance significantly.

 We reduce the mismatch by data augmentation with different simulation strategies. In the testing, the simulated deep speaker embedding features are fused with the original enrollment embedding features.

\subsubsection{Randomly simulation of the enrollment data}

Using \textit{pyroomacoustics}, 5 simulated far-field utterances are randomly generated based on the original enrollment utterance. Then the speaker embedding from the simulated and original utterance are averaged to serve as the enrollment embedding. 

\subsubsection{Testing background aware simulation of the enrollment data}
To further reduce the gap between the close-talking enrollment utterance and the far-field testing speech. We train an ASR acoustic model using AISHELL-2\cite{aishell2_2018} based on GMM to align the 'ni hao, mi ya' test utterances as speech. The utterance of non-speech is background noise. Instead of the random simulation of the enrollment data, we use the background noise of the testing utterance(exactly in the trial) to perform enrollment augmentation.
 
\section{Experiments}
\subsection{Dataset}
\subsubsection{Text-independent corpora}

The AISHELL-2\footnote{\texttt{http://www.aishelltech.com/aishell\_2}} is an open and publicly available Chinese Mandarin speech recognition dataset. In this study, we use the \textit{iOS} channel of the dataset, which contains 984,907 close-talk utterances from 1,997 speakers. We use this dataset to simulate far-field utterances for the pre-training of the deep speaker embedding.

\subsubsection{Text-dependent corpora}

The mandarin wake-up word \textit{'ni hao, mi ya'} was chosen in our experiments. We use \textit{AISHELL-wakeup} data as the fine-tuning training data and \textit{AISHELL-2019b-EVAL} as the test set. Based on our previous experimental results, the last 44 people in AISHELL-2019B-EVAL is more challenging, so we select the utterances of the last 44 people as the test dataset. 

In this paper, we have two tasks, close-talking enrollment task and far-field enrollment task. The testing data for both tasks are far-field utterance. In the case of the close-talking enrollment with far-field testing, we used the data from close-talking HIFI mic for enrollment. In the case of the far-field enrollment with far-field testing, we used data from one microphone array which is 1m away from the speaker for enrollment. 

We proportionally select 1m, 3m and 5m distance array data  as test audio in trials. The keys ratio of target and nontarget is one to one.  
%We fine-tune the pre-trained \textit{ResNet2d} model with six-channels data in each array and the HIFI microphone data. Since the HIFI microphone does not have multiple channels, the \textit{ResNet3d} model can only be fine-tuned with data collected by microphone arrays. 

\subsection{Baseline system and fine-tuned model}

The performances of these models  are shown in Table 2. Comparing the far-field enrollment task with the close-talking enrollment task, the former achieves about 20\% relative improvement in terms of equal error rate (EER). This indicates that although the far-field enrollment data may not be of the best quality, it is able to better match with the far-field testing data. The Basic model (ResNet34-Cosine) in Table \ref{tab:t2} shows the result of the far-field simulated AISHELL2 training data model with cosine similarity scoring. The fine-tune model (ResNet34-FT-Cosine) achieves 20\% improvement comparing with the basic model. The PLDA backend (ResNet34-FT-PLDA) compensates for the channel mismatch with a 20\% relative improvement. For the results ID 1-6, both the enrollment and the testing data are single channel. For systems ID 7-8, each test utterance has all 16 channels from a random selected array. Here we adopt the strategy of speaker embedding level averaging, and the performance is further enhanced. 

\begin{table}[t]
  \caption{EER of different speaker embedding systems. }
  \label{tab:t2}
  \centering
  \begin{tabular}[c]{lllll}
    \toprule
      \textbf{ID} & \textbf{Model} &  \textbf{Enrollment}  & \textbf{EER} \\
      
   \midrule
     $1$ & \multirow{2}*{ResNet34-Cosine} &  far-field  & 6.54\%  \\
     $2$ & &  close-talking & 7.41\% \\
    \midrule
     $3$ & \multirow{2}*{ResNet34-FT-Cosine} &  far-field & 5.08\% \\
     $4$ & &  close-talking &  6.66\% \\
    \midrule
     $5$ & \multirow{2}*{ResNet34-FT-PLDA} &  far-field & 3.92\% \\
     $6$ & &  close-talking &  5.36\% \\
   
   \midrule
     $7$ & {ResNet34-FT-PLDA} &  far-field & \bf{3.7\%}  \\
     $8$ & {+embedding level averaging }&  close-talking &  \bf{4.71\%} \\
     
   \midrule
     \multicolumn{2}{c}{Fusion (1 + 3 + 5 + 7)} &  far-field &  \bf{3.29\%}  \\
     \multicolumn{2}{c}{Fusion (2 + 4 + 6 + 8)} &  close-talking &   \bf{4.02\%} \\
   
     \bottomrule
     \end{tabular}
\end{table}

\subsection{Enrollment data augmentation}

Table \ref{tab:t3} shows the EER of enrollment data augmentation. From the enrollment condition without enrollment augmentation, we observe that the ID 5-6 results of the close-talking enrollment with far-field testing scenarios always have a degraded performance compared to the far-field enrollment with far-field testing scenarios. The main reason is the channel mismatch between the enrollment utterance and the testing utterance. 

The results indicate that the enrollment data augmentation can reduce the gap between the far-field enrollment condition and the close-talking enrollment condition. Also the performance of testing background aware simulation is slightly better than randomly simulation.

\begin{table}[t]
  \caption{EER of enrollment data augmentation }
  \label{tab:t3}
  \centering
  \begin{tabular}[c]{ll}
    \toprule
      \textbf{Enrollment condition} & \textbf{EER} \\
      
   \midrule
	Clean embdd & 5.36\%  \\
	Real far-field embdd & 3.92\% \\
    \midrule
   	Clean embdd + 5 random simulated far-field embdd & 4.73\% \\
   	Clean embdd + 1 test condition aware embdd & 4.59\% \\
     \bottomrule
     \end{tabular}
\end{table}

\subsection{System fusion}
We performed the system fusion at the score level.For the far-field enrollment task, systems with ID 1,3,5,7 are fused. For the close-talking enrollment task, systems with ID 2,4,6,8 are fused. The results show that different system are complementary to each other.

\section{Conclusions}
In this paper, we describes the \textit{HI-MIA} database collected in the far-field acoustic environment. The database contains multi-channel far-field speech data that could be used in different applications, such as text dependent far-field speaker verification, wake-up word detection and speech enhancement. The database has two sub datasets. One named AISHELL-wakeup could be used as the training data and the other named AISHELL-2019B-eval is designed as the development and testing data. Besides, we develop several baseline systems under the close-talking enrollment with far-field testing protocol. We also introduce fine-tuning method for training with limited text-dependent data. Enrollment data augmentation strategies is also investigated to reduce the mismatch between clean enrollment utterance and far0field testing utterances. Results show that augmenting the enrollment utterance towards the acoustic environment of the test utterance can effectively improve system performance. 

\label{sec:conclusions}

\section{Acknowledgements}
This research is funded in part by the National Natural Science Foundation of China (61773413), Key Research and Development Program of Jiangsu Province (BE2019054), Six talent peaks project in Jiangsu Province (JY-074), Guangzhou Municipal People's Livelihood Science and Technology Plan (201903010040), Science and Technology Program of Kunshan City.

\bibliographystyle{IEEEtran}

\bibliography{arxiv.bib}

\end{document}